\documentclass[11pt, a4paper]{article}
\usepackage[utf8]{inputenc}
\usepackage[margin=33mm]{geometry}
\usepackage{graphicx}
\usepackage{authblk}
\usepackage{amsmath}
\usepackage{natbib} 
\setcitestyle{authoryear,open={(},close={)}} %Citation-related commands

\title{Modelling the multi-scalar effect of commuting on exposure to diversity }
\author[1,*]{Valentina Marin}
\author[1]{Carlos Molinero}
\author[1]{Elsa Arcaute}
\affil[1]{Centre for Advanced Spatial Analysis (CASA) University College London, UK}
\affil[*]{corresponding author: v.marin@ucl.ac.uk}
\date{}

\begin{document}

\maketitle

\begin{abstract}

Urban systems are primarily relational. The uneven intensities and distribution of flows between systems of cities results in hierarchically organised complex networks of urban exchange. Distinct urban spatial structures reflect the diversity of functional and social patterns which vary or remain constant across multiple scales. In this work, we examine the impact of commuting on the potential accessibility to spatial and social diversity, and the scalar relations that may exist. We first define relational scales by conducting a process of percolation on the commuting network, as a hierarchical clustering algorithm. This gives rise to a nested structure of urban clusters based on flow connectivity. For each cluster at each scale, we compute measures of commuting structural and social diversity by examining the spatial distribution of origin-destination pairs, and the distribution of workers' skills and occupations. To do this, we make use of global entropy measures allowing us to quantitatively analyse the reachable diversity across scales. Applying this methodology to Chile, we observe that the hierarchical accessibility to the wider system of cities and the patterns of spatial interaction, significantly influence the degree of exposure to diversity within urban systems. This framework examines connectivity-based diversity at multiple scales, and allows for the classification of cities and systems of cities according to the spatial and social dimensions of commuting dispersion. Such insights could contribute to the planning of infrastructural projects connecting the urban system at different scales, while also guiding a strategic relocation and redistribution of economic activities at regional levels.

\end{abstract}

\section{Introduction}

Geographic space is not uniform and presents great heterogeneity and spatial diversity. The elements that make up urban systems, as well as the relationships that keep them interconnected are unevenly distributed across different locations and scales \citep{batty1994fractal,Bretagnolle2006, Pumain2006,Halas2014}. This results in urban spatial structures \citep{Sohn2005, Fusco2011} with different patterns of spatial arrangements and interactions, and varied levels of connectivity and accessibility, that shape social and economic behaviour with the possibility of instigating social disparities \citep{MaturanaMiranda2012,Burger2014}. Since urban systems are primarily relational in nature, the patterns of spatial interactions are essential for understanding the complex networks of interdependence and differentiation between different cities \citep{batty2018inventing}. In this context, the study of flows and exchange between spatial locations can provide valuable knowledge on urban relationships. Flows can take many different forms such as trade, information, social or commuting networks, each with their own unique structure of organisation, offering a range of insights into the patterns observed in urban areas \citep{Burger2014, Hesse2011,Fusco2011}. Within the various functional relationships, commuting networks are particularly important. Their consistent and representative patterns provide a comprehensive understanding of the overall interactions within and between cities \citep{Hanson1988,CLARK2003199}. Multiple research efforts have been dedicated to analysing the patterns of home-to-work links, to explain variations in economic growth \citep{Goetz2010}, to determine the limits of functional regions \citep{Kropp2014}, to examine socio-demographic characteristics \citep{Montis2007, Mare2019}, to categorise cities by their commuting patterns \citep{Louail2015}, to investigate the changes in mobility over time \citep{Patuelli2007,DeMontis2011,Louail2014}, or to study the transmission of infectious diseases \citep{Balcan2009} to name a few. The study of commuting networks also involve different methods such as: the examination of the networks densities given by the magnitude of inbound and outbound flows, the examination of trip distances and times to understand work-related travel costs, the analysis of network connectivity to identify key locations and their roles, or the assessment of the spatial distribution of commuting flows given by housing and employment opportunities. In the context of this paper, we are particularly interested in analysing the distribution of commuting flows by means of diversity measures. Diversity  provides valuable information about the spatial and social organisational structure of urban systems \citep{Lowe1998,Cheng2013,Pappalardo2016, Marin2022}. From a spatial perspective, a diverse commuting network leads to multiple potential interactions between cities, resulting in increased  accessibility and a more integrated and adaptable urban system \citep{Reggiani2010}. From a social perspective, labour diversity for instance, can provide a range of different skills and resources, which could improve the overall performance and productivity of the system.

By analysing the diversity of flows among origin-destination pairs, key patterns of dispersion or concentration can be identified, providing insight into the various typologies of relationships, as well as the relative importance of urban centres and their roles in the overall functioning of the urban system \citep{Goetz2010,Burger2014}. High concentrations of flows in certain destinations can reveal underlying spatial disparities and high dependency, which can make the entire system overly reliant on few urban centres. In contrast, diverse origin-destination flows within an urban region can help distribute the benefits and burdens of economic development and ensure more equitable access to opportunities, resources and services. Commuting is also related to social diversity by potentially increasing workers' exposure to difference \citep{Mare2019}. By analysing the distribution of the active population based on their skills for instance, a better understanding of the uneven spatial distribution of different population subgroups can be gained. This is especially relevant to study socio-spatial disparities in regional performance. One key aspect of workers-skills diversity is the potential for spillovers of knowledge and ideas to occur as individuals come into contact with a wider range of people and places and feed back into their home areas \citep{Goetz2010}. This can contribute to increased understanding and collaboration among people from different backgrounds, fostering innovation as well as increased access to economic mobility.

\subsection{Scalar variation}

Spatial and social diversity patterns are scale-dependent, meaning they can vary greatly depending on the scale of observation \citep{ Wu2004}. This is because urban dynamics are the result of the interplay of various phenomena operating at different scales. The organisation of elements within a specific geographic location is influenced by local and internal processes as well as external processes operating at larger regional scales \citep{Maciejewski2016, Rozenblat2022}. Different phenomena may have different relevant scales, which in turn means that there is no unique scale for describing spatial or social diversity. \citep{DeMontis2011, Boussauw2012}. A multi-scalar approach to the structural organisation of urban systems provide a more comprehensive view of how patterns evolve across different scales.

The scalar variation of the patterns of flow diversity in commuting networks could be influenced by many factors, such as governmental policies, transportation infrastructure, or the distribution of specific industries and populations, leading to different commuting patterns across scales. For example, commuting can become more diverse when observing the system at larger scales. At a local level, a city may serve as the main urban centre in a monocentric area while at a larger scale, it can also engage with other urban centres that specialise in distinct functions, resulting in a more diverse and polycentric interaction within the region. On the other hand, commuting patterns could also remain relatively stable across scales, or change into more concentrated patterns instead. In the latter case, a city may have similar interactions with multiple cities in a small polycentric area, but be embedded in a larger monocentric region where a dominant economic centre attracts most intra-regional flows \citep{Lowe1998}.

In terms of social organisation, variation of patterns in workers skills and occupations at different scales, may be influenced by personal choices and demographic characteristics, but also by specific urban spatial interaction patterns. At smaller scales, a diverse economy within a city could lead to a wider range of job opportunities and a diverse array skills. In contrast, homogeneous industries at a local level result in a more uniform skillset. As the scale increases, patterns could shift based on the degree of interdependence among cities and their interaction structures, as well as the level of specialisation of the other cities within the same region. In regions where cities have complementary functions and robust exchange, there is often a higher diversity of skills among workers. In contrast, in regions where cities have similar functions there may be a tendency for certain industries to dominate and for workers to specialise in a narrow range of skills. 

Understanding scalar variation both upwards and downwards is crucial for decision-making processes in areas such as transportation planning, economic development, and equitable distribution of resources and opportunities. Governance must consider the varying organisational scales and adjust accordingly by embracing a multi-scalar perspective and a more flexible framework \citep{Anderies2006,Cumming2006}. Monocentric organisations evolving into more diverse regional structures, would require greater collaboration between local administrative units and other entities, through the creation of mutual agreements, shared facilities, and joint regional programs. In contrast, as systems become more centralised at larger scales, effective governance is crucial in managing high levels of dependency, balancing the benefits of centralisation with the drawbacks of potential risks of reduced diversity and increased vulnerability.

\subsection{Defining scales}

Defining relevant scales is essential to studying scalar variation in patterns of urban interaction. The scales of research can impact our understanding of geographical processes \citep{Jonas2011,Rozenblat2013Conclusion}, then, establishing appropriate spatial scales that aligns with the ones of the phenomena is crucial and presents a methodological challenge \citep{Montello2001, Allen2014, Walz2015}. Urban scales are usually defined by government agencies for administrative, economic, and political purposes. However, these traditional definitions may not accurately reflect the scales of urban processes lacking practical significance. Instead, a more accurate representation of an urban scale can be identified by examining the patterns of interactions between different spatial units, such as through mobility or socioeconomic relations \citep{Fusco2011,Yin2017}. Rather than viewing cities as isolated entities, this approach allows cities to be understood as embedded within larger networks of cities in a national urban hierarchy \citep{batty1994fractal,Bretagnolle2006}. As individuals move throughout a territory, certain areas become more interconnected than others. In this context, \emph{scale} refers to the spatial extension of processes given by the intensity of these interactions. By arranging these interactions into successive scales based on their magnitude, a hierarchical organisation can be discerned, serving as a basis for multi-scale analysis that acknowledges changes in underlying patterns as scales shift.

This paper examines the uneven distribution of commuting flows across scales, resulting in distinct urban structures with varying degrees of exposure to social and spatial diversity. We apply a percolation process on the commuting network to identify relevant scales by constructing the hierarchical structure of urban clusters. Then we compute the cophenetic distance on the hierarchical structure based on the intensity of relationships between cities in the system. This allows us to investigate how the scalar accessibility between clusters, may be related to diversity patterns. For each cluster at each scale, we compute diversity by examining the spatial distribution of origin-destination pairs as well as the range of  workers' skills and occupations. We use Shannon's entropy and a global entropy metric introduced in \citep{Marin2022}. Finally, we perform a multivariate analysis using all multi-scalar outputs to identify clusters of closely related systems of cities. Through this process, we are able to determine structural categories of networks and classify cities and systems of cities based on the spatial and social dimensions of commuting dispersion.

\section{Methods}
 
In this study, we examine observed transitions from one scalar pattern to another with regards to the diversity measures computed to examine spatial and social changes in commuting networks. The paper implements a methodological framework consisting of three key stages. First, we aim to retrieve the hierarchical organisation of commuting patterns to define appropriate scales for analysis. Then, we analyse diversity patterns of commuting clusters at different scales, taking into account the spatial distribution of flows and skills levels. Finally, we identify categories of systems of cities based on their multi-scalar accessibility to diversity, given by the outputs obtained in the previous measures.

Chile's unique geography and economic composition provide an interesting case study for exploring patterns in commuting networks. The long and narrow shape of the country, coupled with its varied climate, has resulted in diverse economic specialisations across different cities, leading to distinct commuting patterns and potential for varied spatial interaction structures. Studying the commuting networks in Chile can provide insights into how these interactions shape skill diversity and inform the distribution of economic activities. Additionally, the level of connectivity between cities varies across the country, with some cities being highly interconnected while others are relatively isolated. The concentration of economic activity in the capital city of Santiago poses significant challenges in achieving equitable and sustainable spatial development across the country.

\subsection{Commuting network} 

To construct the most representative Chilean commuting network we use the 2002 Chilean Census origin-destination matrix, which is the latest census edition containing origin-destination micro-data. We compared the census dataset with the one extracted from the National Employment Survey (ENE), a more recent survey with a statistically representative sample for the years 2015 to 2019. We found that there was a correspondence of $94\%$ with respect to Census flows origin-destination patterns. The advantage of using Census data is that it contains more detailed information about workers' skills and a much larger sample. It includes information on the skill level required for elementary occupations as defined by the International Standard Classification of Occupations (ISCO), which is used for this analysis (Table \ref{isco}). The data is aggregated at the commune level and then mapped to the city level, forming a network where cities are represented as nodes $v$ and the flows of commuters between cities represented as directed and weighted links $m$. The dataset removes commuting flows that exceed a 3-hour driving duration to provide a more precise assessment of the strength and consistency of inter-city connections, considering that daily commuting reflects the typical and repetitive movements between cities. The limit of 3 hours for driving commuting time across the country is established by the CASEN (National Socioeconomic Characterisation Survey of Chile).

\begin{table}
\centering
\makebox[\textwidth][c]{\includegraphics[width=0.75\textwidth]{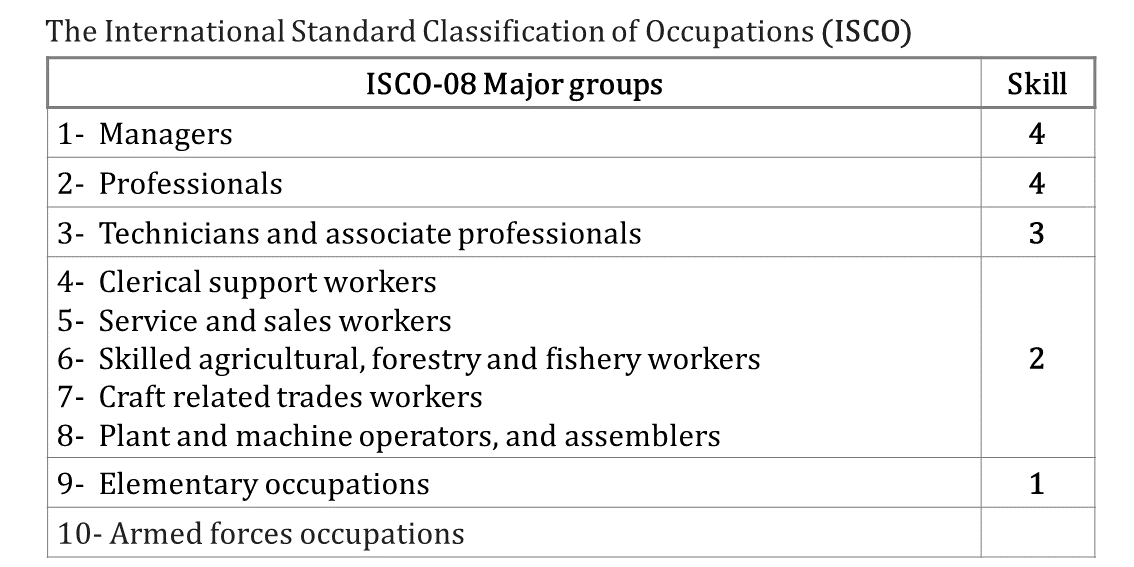}}
\caption{\textit{\textbf{Distribution of occupations and skills based on the International Standard Classification of Occupations (ISCO) system. } The table categorises the skill levels for different ISCO categories, ranging from basic skills (1)  to expert-level skills (4). Skill levels categorise the level of knowledge and expertise required for an occupation, based on the complexity and specialisation of the work involved, as well as the degree of autonomy and responsibility required.}}
\label{isco}
\end{table}

\subsection{Commuting hierarchical organisation}

We use percolation analysis to retrieve the scalar structure of Chilean commuting network. Percolation theory is usually used for analysing phase transitions in connected clusters, which can be used to understand the flow of information within a system \citep{Gallos2011}. It examines the global structure of connectivity by starting with disconnected elements and gradually merging them based on a thresholding process until the largest cluster encompasses the entire system. This approach has been applied in various fields, including functional brain networks \citep{Gallos2011}, disease spread \citep{Newman2002, Gallos2012}, and urban systems \citep{Rozenfeld2008, Rozenfeld2011, Fluschnik2016, Arcaute2016, Piovani2017, Behnisch2019, Sarkar2020}. In this study, we implement a similar threshold procedure as in prior research \citep{Arcaute2016,Piovani2017}, but applied to a commuting network. The procedure involves obtaining sub-graphs of connected links where the weight $w_{ij}$ is above a set threshold $\tau$, meaning $w_{ij} \geq \tau$. The link weights in commuting networks are typically determined by the overall outflow count \citep{Rabino1997}. However, this paper presents an alternative approach of defining link weights through a dependence parameter. The parameter is computed in terms of the magnitude of the flow directed towards a destination in relation to the total outflows in the origin. This approach allows clusters to be induced not only by the intensity given by the number of flows, but also by the relative dependency of the interrelationships holding them together. In order to clarify this, we define a \textit{dependence} parameter between two systems based on the out-flows.

\begin{itemize}
\item Dependence-based weight:
\begin{equation}
  w_{ij}= \frac{t_{ij}}{\sum_{j}  t_{ij}}
\end{equation} 
where $ t_{ij}$ is the out-flow from node $v_{i}$ to $v_{j}$ and $\sum_{j}  t_{ij}$ is the total out-flow from the origin $v_{i}$ to all possible destinations.
\end{itemize}

In essence, the dependence parameter aims to quantify the degree to which an origin node relies on or is connected to a specific destination node. A high dependence parameter for a destination indicates that a significant fraction of commuters from the origin commute to that destination, suggesting that the destination holds a relatively high importance for that particular location. By using the dependence parameter instead of focusing solely on the total volume of flows between cities, we gain a more comprehensive understanding of the relative importance of connections between cities, regardless of their size. When only the total volume of flows is considered, larger cities naturally tend to exhibit higher strength in their connections, as they have larger population and therefore generate larger commuter flows. However, this approach may overlook significant relationships between smaller cities, whose absolute flows are lower in comparison.

\begin{figure}[h!]
\centering
\makebox[\textwidth][c]{\includegraphics[width=1\textwidth]{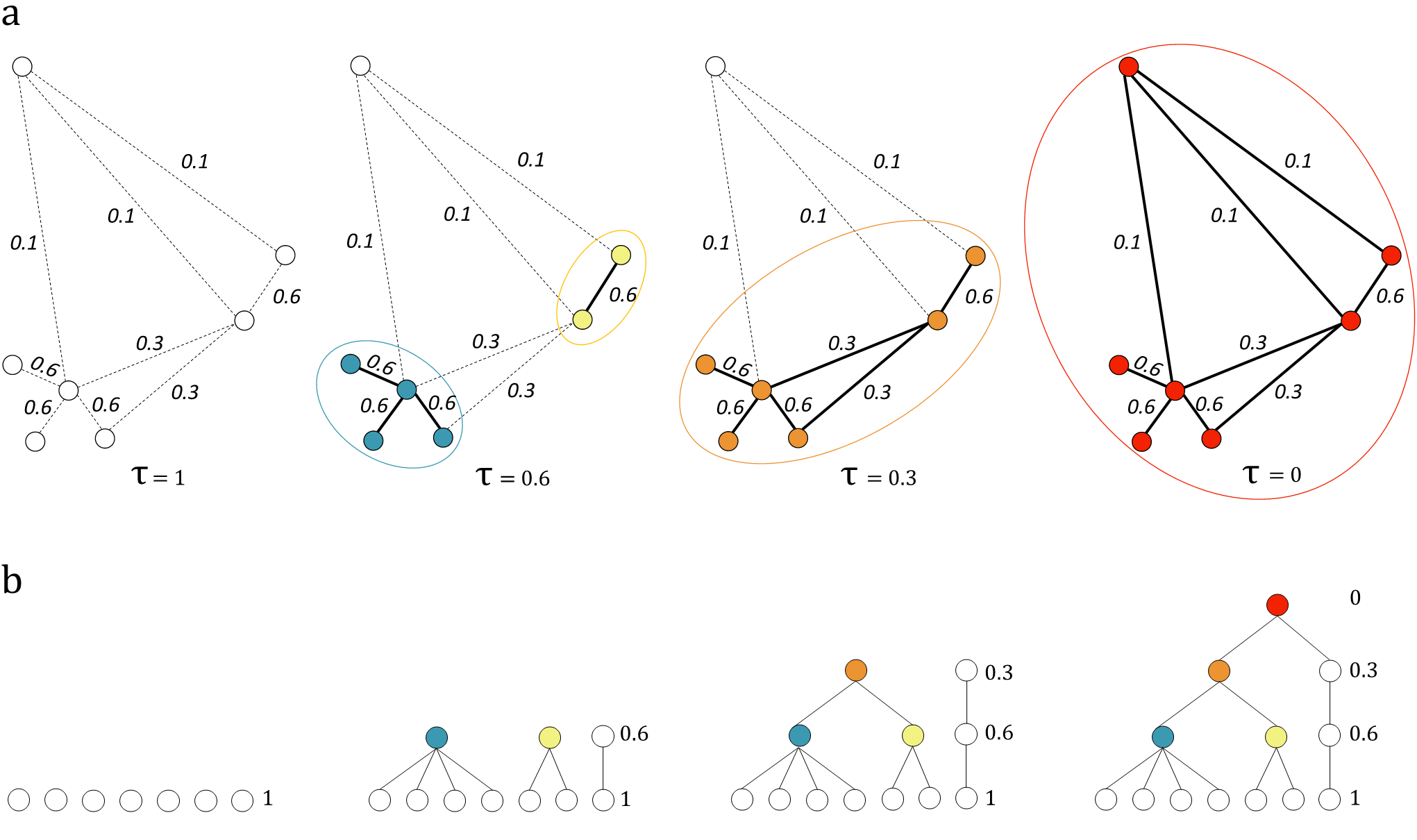}}
\caption{\textit{\textbf{Hierarchical percolation process:}} a) Toy example of the percolation process in the network, illustrating cluster formation at various thresholds defined by the dependence parameter. b) Iterative construction of the dendrogram following cluster formation, where the maximum threshold $\tau=1$ corresponds to the terminal nodes of the structure, and the minimum $\tau =0$ to the top, where all elements are clustered together.}
\label{process}
\end{figure}

The percolation procedure starts from the threshold $\tau= 1$, by which all existing links are disconnected, since the maximum dependence weight $w_{ij}=1$ is not satisfied in any case. A maximum dependency would suggest that all commuting individuals within a specific area are required to commute exclusively to one location. As the threshold decreases, a sequence of percolation transitions takes place, resulting in clusters of links connected by $w_{ij} \geq \tau$. The clusters formed at the initial thresholds exhibit significant interdependence among their components, which is primarily determined by a higher dependency parameter rather than a larger count of flows within the network. A hierarchical structure can be reconstructed by arranging the clusters obtained from percolation. This is because the clusters are organised in a nested manner, where those generated with a threshold $\tau_1$ are contained within those generated with a following threshold $\tau_2$ (Figure \ref{process}). Significant scale transitions can be identified when there are sudden changes in the size of the giant cluster as the thresholds change \citep{Arcaute2016}. These discontinuities do not represent actual breaks in the structure, in this case they rather reflect changes in the strength of interdependencies \citep{Simon1962,Holling2001UnderstandingSystems}. The thresholds where these changes occur are considered as relevant scales, that are arranged vertically in a dendrogram structure, with the first thresholds at the bottom (high dependence) and the final thresholds at the top (low dependence).

\subsection{Hierarchical distance}

Geographical distance and accessibility play a crucial role in spatial dynamics, but as argued by \citep{Rozenblat2013}, other distances also prove to be important. Measures of proximity between urban areas that don't rely exclusively on physical proximity, provide complementary perspectives, expanding our understanding of systems interactions \citep{Rozenblat2022}. In this paper, distance refers to a relational measure that assesses the relationship intensity between cities. Scalar accessibility relates to the ease of accessing other cities based on its position in the hierarchical structure determined by its dependencies on each system component. The paper uses the Cophenetic distance $D$, a pair-wise distance similarity measure, to calculate scalar accessibility. It is defined as the threshold at which two terminal clusters $n_i$ and $n_j$ (cities in this case) are joined together in the same cluster \citep{Sokal1962}. $D$ is computed for every pair of cities at the terminal level of the dendrogram. A larger value of $D$ means a higher scalar distance between a pair of cities and a lower potential for inter-dependencies between them.

\subsection{Commuting and skills diversity}

We examine diversity in commuting networks from two approaches, one that looks at the distribution of workers skills, and other that looks at the spatial distribution of origin-destination flows. For the first one, we use the standard Shannon's entropy measure, which from a more ecological approach considers the richness and evenness of different skills. The more types of skills present (richness) and the more equally abundant they are (evenness), the less predictable the distribution of skills \citep{Hamilton2005,heip1998indices}. Thus, when used as a measure of diversity, greater uncertainty indicates higher diversity. In the second part, we use a measure derived from Shannon Entropy which is based on an information-theoretic approach to networks \citep{Sole2004, Wilhelm2007, Marin2022}. Diversity in this context looks at the heterogeneity of the arrangement of links and nodes in the network of flows. We compute both normalised measures for each cluster at each scale of the dendrogram, allowing for the comparison of clusters of varying sizes and the assessment of changes in the patterns as the scale evolves.

\textbf{Skills diversity}: We use Shannon entropy to analyse the distribution of workers' skills within each cluster according to the number of skill categories and its relative proportion. The entropy outputs can be interpreted as the level of diversity in the distribution of skills. A high entropy value represents a diverse and evenly distributed mix of skills, without predominant skill sets. On the other hand, low entropy values show non-diverse and imbalanced distribution of skills, where certain types are over or under represented. Workers' skills are defined by the ISCO classification which is a standard description of jobs and occupations, categorising them into skill levels based on required education, training, and experience. 

\begin{itemize}
\item Global Skills entropy \citep{Shannon1948}:
\begin{equation}
  H_{S}= -\sum_{i}  p_{i}  \log p_{i}
\end{equation} 
where $ p_{i}$ is the probability of workers with a particular skill $i$ in the urban area. To make this measure comparable between urban systems of different sizes we normalise it by the possible maximum entropy value for a system of that size. in this case this is obtained by the log of the total number of workers ($T$). Then the normalise skill entropy will be given by $H_{S}\log(T)$
\end{itemize}

\textbf{Commuting spatial diversity}: The diversity of origin-destination patterns in urban commuting networks is analysed through the computation of global in-commuting entropy described in \citep{Marin2022}. This information-theoretic approach provides insight into the distribution of flows in the network, and the centralisation and dispersion of urban spatial structure. The global entropy approach considers the interdependence of all network components, where changes in the local patterns impact the overall network entropy. This measure examines the probability of distribution of in-commuting flows across all urban units. A skewed probability distribution of flows results in reduced entropy, characterised by a monocentric pattern due to the concentration of flows in a few dominant areas. On the other hand, an evenly distributed probability results in higher entropy and uncertainty, exhibiting a polycentric pattern with a more balanced distribution of importance among urban centres.

\begin{itemize}
\item Global in-flow entropy at node level \citep{Marin2022}:
\begin{equation}
  H_{C}= -\sum_{\forall j} \left(\sum_{\forall i} p_{ij}\right)  \log  \left(\sum_{\forall i} p_{ij}\right)
\end{equation} 
where $\sum_{i} p_{ij}$ is the probability of in-flow to node $v_{j}$, considering the sum of all flows arriving to $v_{j}$ from every possible node $v_{i}$. To normalise this metric and make it comparable between systems of different sizes across scales, we use the log of the maximum possible total number of nodes in a system ($N$), then $H_{C}/\log(N)$.
\end{itemize}

\subsection{Multi-dimensional analysis}

To explore the relation between the different diversity metrics and evaluate their effect on changes in patterns across multiple scales, we use dimensional reduction techniques on the outputs obtained from these metrics at various thresholds. To do this, we employ Principal Component Analysis (PCA) that transforms a set of possibly correlated variables into a set of linearly uncorrelated variables called principal components \citep{pearson1901}. It simplifies the data by reducing its dimensions while preserving significant information with the first principal component retaining the most information and variance. By employing clustering algorithms, such as k-means, we identified different groups of categories based on the patterns captured by the principal components. This enabled us to categorise cities into meaningful groups based on diversity variance at different scales.

\section{Results}

In the following section, we applied the methodology to the Chilean commuting network. The results revealed a significant variation in spatial interaction structures of city systems throughout the territory. This variation is assessed based on the spatial distribution of flows, the workers' skillsets and their scalar distances given by different levels of commuting dependency.

\subsection{The scalar organisation of the system's interdependencies}

\begin{figure}[h!]
\centering
\makebox[\textwidth][c]{\includegraphics[width=1\textwidth]{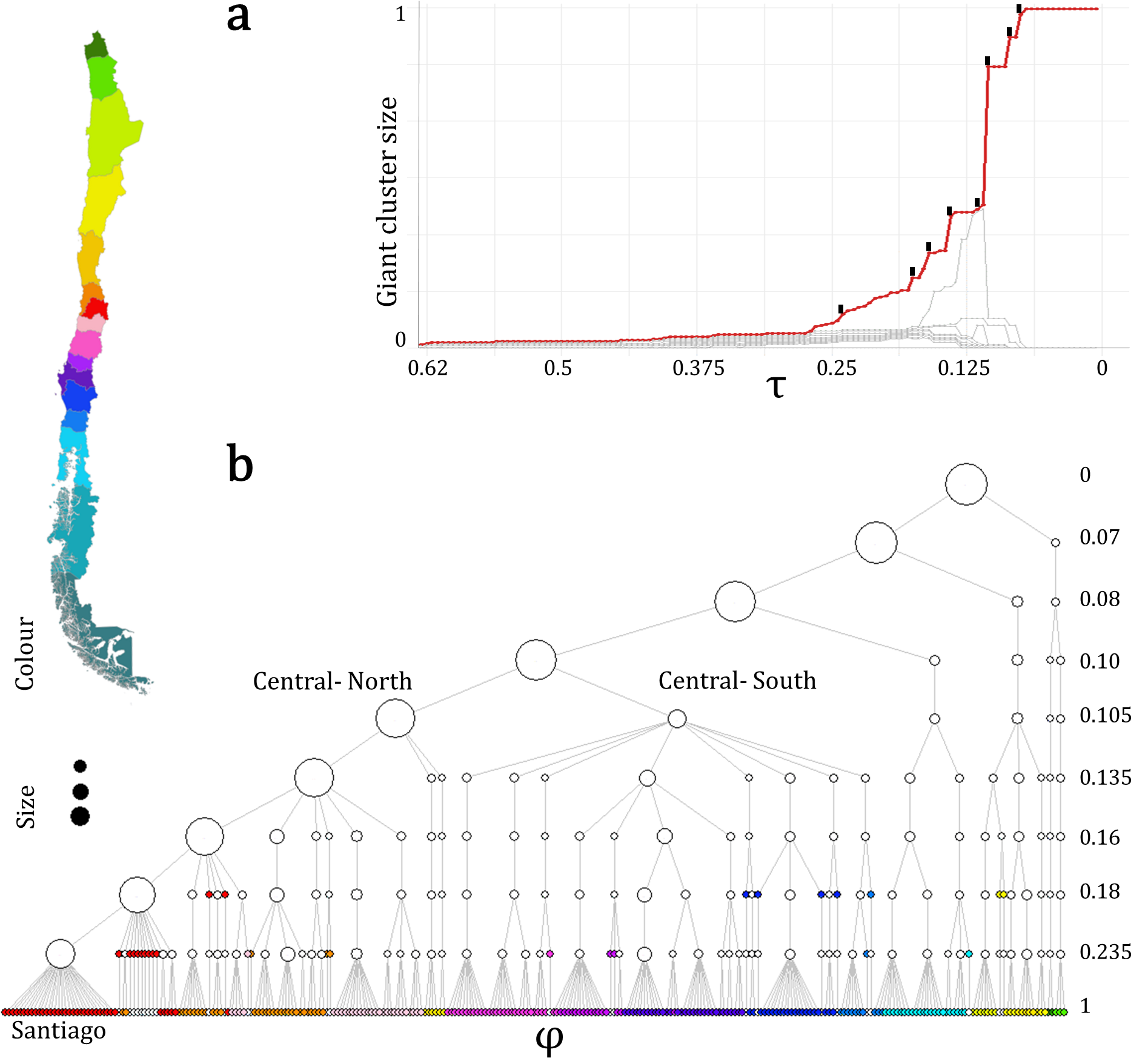}}
\caption{\textit{\textbf{Hierarchical organisation of Chilean commuting network.} } (a) The plot shows the size of the giant cluster as the percolation threshold increases. The x-axis represents the threshold value while the y-axis shows the size of the giant cluster. (b) The dendrogram in the figure represents the resulting clusters at relevant thresholds, with colours indicating the regions for each city and size indicating the total flow.}
\label{perco}
\end{figure}

The heterogeneities of the interactions between the different areas of the urban system is what creates the regional divisions and more generally, the observed organisation of the space. The latter can be extracted by observing which areas interact at varying threshold levels. Starting at the lowest level, one can think of the system as being fully disconnected. As the threshold moves from $\tau=1$ to $\tau=0$, areas start to merge, until forming a unified cluster that spans the entire territory. During this process, the system undergoes a series of continuous phase transitions. The multiplicity of transitions is a signature of a hierarchical organisation. These are traditionally studied by looking at the sudden changes in size of the giant connected component (Figure \ref{perco}a). We will call \emph{critical thresholds}, the thresholds at which these transitions are observed. These critical thresholds are then used to construct the dendrogram that displays the relevant clusters in the relationships between cities and the rest of the territory (Figure \ref{perco}b). 

The dendrogram structure offers a visual representation of the network's connectedness at various scales. Clusters are formed based on commuting flows between cities, which reflects a degree of dependency for they functioning. In Figure \ref{perco}b, cities are at the bottom of the structure representing the terminal clusters $n_i$ of the dendrogram. We can observe that small systems of cities are formed when the threshold is at $\tau = 0.235$. From this threshold, the systems of cities display greater variability, with provincial, regional and other larger forms of organisation clustering at different thresholds along the country. Cities in the central region of the country, in red and orange, are clustered in the early thresholds ($\tau = 0.18 - 0.135$), indicating a strong interdependence. Conversely, the cities on the peripheries, in green, yellow and cyan, exhibit a slower formation of clusters, joining at final thresholds ($\tau = 0.10 - 0.07$), showing a relatively weak inter-urban relationship.

 \subsection{Scalar Accessibility}

The cophenetic distance $D$ between cities in the dendrogram represents their relative scalar positions given by the strength of urban relationships. A smaller distance between a pair of cities indicates a stronger interdependence, while a larger distance indicates a weaker interdependence. 

\begin{figure}[h!]
\centering
\makebox[\textwidth][c]{\includegraphics[width=1.1\textwidth]{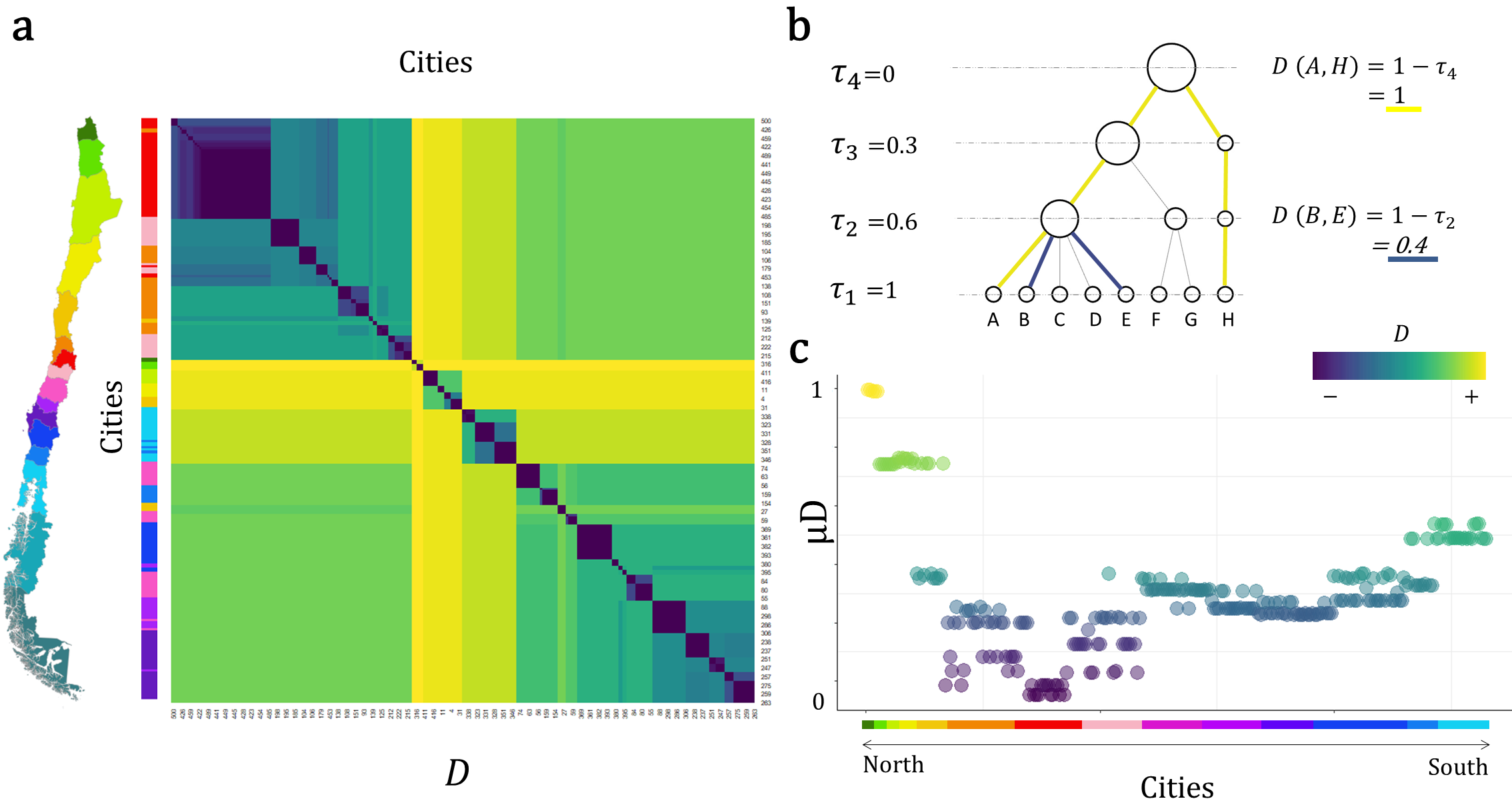}}
\caption{\textit{\textbf{Scalar accessibility given the cophenetic distance ($D$) between cities in the commuting network.} (a) The cophenetic distance matrix between every city. (b) Toy example of cophenetic distances in a dendrogram.(c) Mean cophenetic distance of each city, where the x-axis represents the cities ordered north to south, and the y-axis represents the mean cophenetic value.}}
\label{cophe}
\end{figure}

Figure \ref{cophe}a, shows the matrix of hierarchical distances between each city and all the cities in the system. The diagonal of the matrix displays the smaller hierarchical distances, where cities  with highest interdependence can be identified. A strongly connected system consisting of cities in the central-north of the country can be seen in the upper left corner of the matrix. In the opposite corner, another system made up of cities in the central-south can also be recognised, but with greater distances than the previous case. These two agglomerations can be identified in the dendrogram $\varphi$ shown in Figure \ref{perco}b, at a threshold value of $\tau=0.105$. The cities located at the extremes of the country have low scalar distances with only a few cities, while being considerably distant from the majority of the other cities in the system. Figure \ref{cophe}c, shows the mean of the cophenetic distances for each city, ordered from north to south. The cities in the centre of the country have a lower mean scalar distance, which suggests that they have greater interdependence with the rest of the country. This reflect a high level of centralisation in the system which have important implications for their role in the overall economic and social interactions. This important employment hub is centred around Santiago, the capital, and functions as a crucial node in the network of regional interactions, underpinning the functioning of the country's economy, and playing an important role in facilitating trade and commerce, in addition to connecting people to jobs and opportunities, and supporting regional and national economic growth. High interdependence means that changes in these cities, whether due to natural disasters, economic fluctuations, or other factors, can have far-reaching effects on other regions and industries. In contrast, the mean hierarchical distance increases towards the more peripheral and geographically isolated cities. This suggests a lower level of integration with the national labour market, which could limit their ability to diversify their economy and access external resources. At the same time, the low interdependence with the wider system, could provide some isolation from external labour market changes. 

\subsection{Scalar exposure to diversity}

This section presents the outcomes of the analysis conducted on commuting and skills diversity. Specifically, the different entropy metrics for each system are assessed at relevant thresholds to explore the potential variation on systems exposure to diversity across different scales.

\subsubsection{Commuting diversity $H_{C}$}

We examine the spatial aspects of commuting concentration and dispersion to identify underlying scalar organisational patterns. Low entropy, reflects low diversity in flow dispersion, with most flows concentrated in a dominant area, following a monocentric structure. In contrast, a high entropy reveals a diverse distribution of commuting flows, with a more balance importance of cities, following a polycentric logic.

\begin{figure}[h!]
\centering
\makebox[\textwidth][c]{\includegraphics[width=1\textwidth]{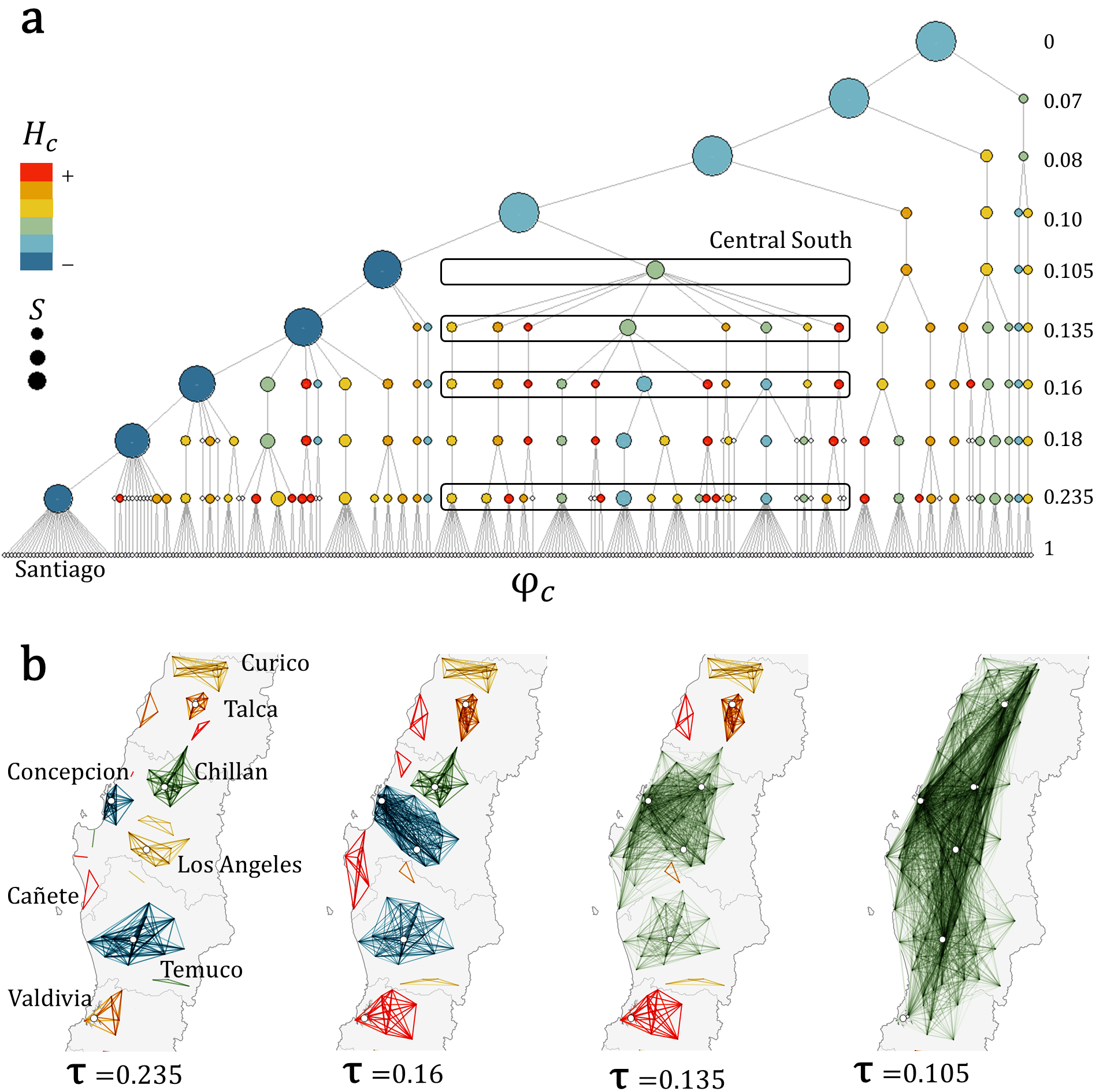}}
\caption{\textit{\textbf{Commuting diversity $H_{C}$ across scales}. (a) $H_{C}$ for each cluster in the dendrogram, with colours indicating the ranking of the values, from blue (lowest value) to red (highest value). The size of each circle represents the flows in each cluster. (b) Commuting network in the central zone of the country at different thresholds, corresponding to the boxed clusters highlighted in the dendrogram. These plots offer a visual representation of the commuting flows within each cluster and the changes in the network structure as the threshold increases.}}
\label{com}
\end{figure}

Figure \ref{com} shows the diversity resulting of the $H_{C}$ measure at each commuting cluster. Findings highlight the different patterns of commuting diversity across scales. Some systems tend to concentrate their commuting patterns as they merged with other clusters, resulting in decreased diversity at larger scales. Conversely, other systems became more decentralised, displaying increased diversity. There are other systems with constant levels of commuting diversity with little to no change in patterns across scales. In Figure~\ref{com} $a$) the main cluster, which initially includes Santiago and gradually assimilates the rest of the clusters at following thresholds, is typically a highly homogeneous and centralised system. It begins with very low diversity and a monocentric structure centred on the capital, and after clustering with the Central-south in $\tau =0.105$, the pattern becomes slightly more diverse but still maintains a low entropy. Despite this, this large system is made up of smaller ones with different diversity values. Figure~\ref{com}b displays the clusters of commuting networks in the central- south area of the country. At $\tau = 0.105$, the central-southern system exhibits moderate diversity and is comprised of city systems that differ in their spatial structures given the diversity of flows. For example, the city system with Los Angeles at $\tau = 0.235$ has high diversity and a polycentric structure, but its diversity decreases on $\tau = 0.16$ after merging with the system containing Concepci\'on. It slightly increases again at $\tau = 0.135$ upon combining with the system containing Chill\' an. Other systems behave differently, the cluster containing  Valdivia, for example, becomes more diverse as the scale increases, whereas the diversity values of the cluster with Curico remain constant across scales until it merges with the other cities from the central-southern region.

The diversity of the spatial interaction patterns pose different challenges for planning and governance. For instance, in a monocentric organisation, policies may need to balance growth and the dependency on the city centre together with the development of surrounding areas. In contrast, in a polycentric system, policies may need to promote collaboration and coordination among multiple activity centres. Changes in these patterns can create challenges at different scales, demanding a coordinated effort for decision-making.

\subsubsection{Skills diversity $H_{S}$}

Measuring the entropy of skills across different levels of organisation can provide valuable understanding of how urban clusters develop and their impact on the diversity of the workforce patterns. High entropy values indicate a diverse range of skills among workers in the resulting clusters, while low entropy values suggest a less varied and uneven distribution of skills. When clusters merge and their patterns of skill diversity are high or complementary, the resulting larger cluster has the potential to increase the overall skill diversity. In these cases, workers with diverse skills and educational levels are brought together, which can even out the proportion of skills within the cluster. As a result, the merged entity benefits from a more balanced and comprehensive skill-set. In the event that workers in one cluster possess skills that are redundant with those in another cluster, the resulting cluster may have a reduced level of overall diversity due to an uneven distribution of skills. This scenario may limit opportunities for workers to interact with individuals who possess different skill-sets.

\begin{figure}[h!]
\centering
\makebox[\textwidth][c]{\includegraphics[width=1\textwidth]{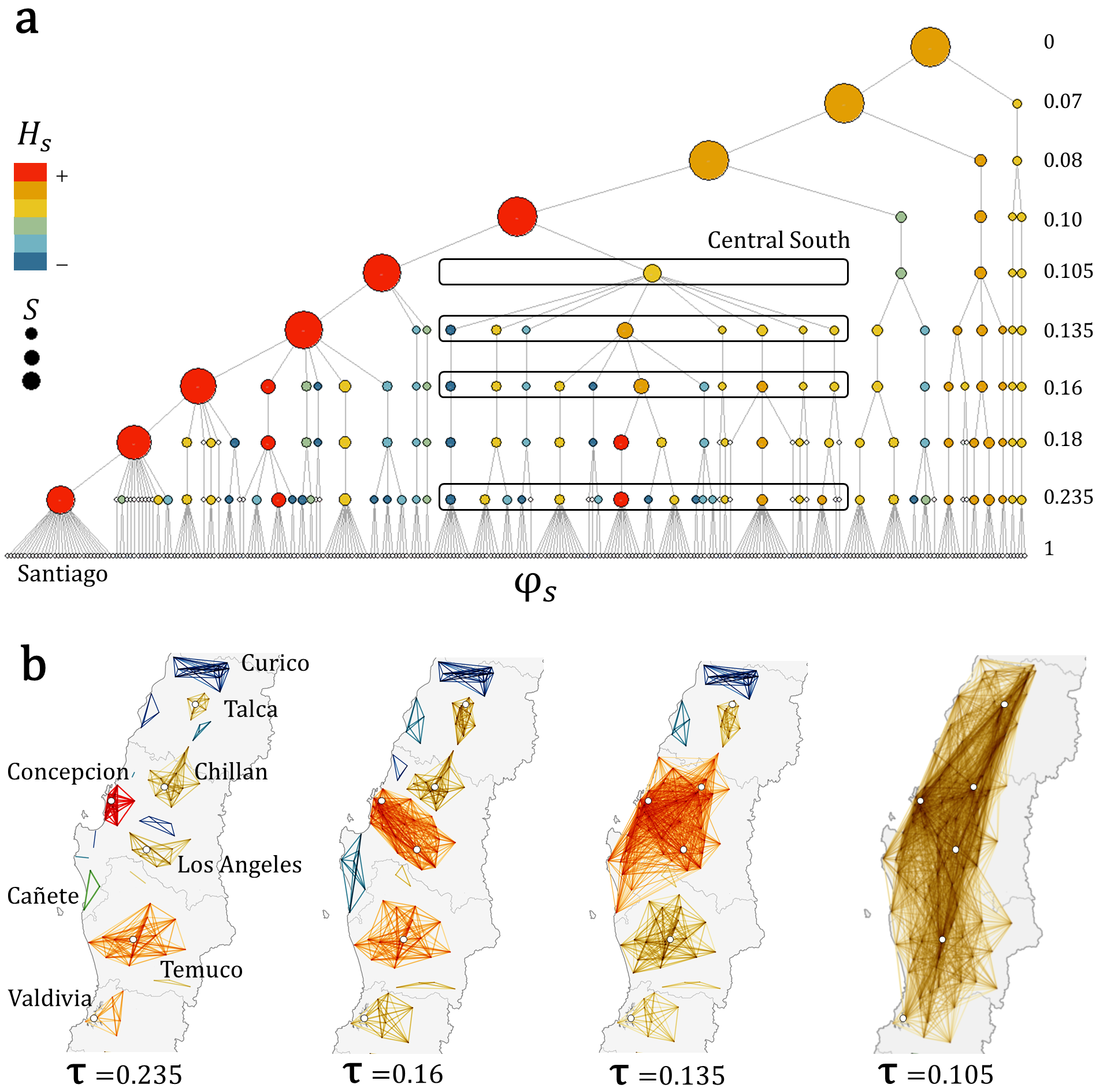}}
\caption{\textit{\textbf{Skills diversity $H_{S}$ across scales} (a) $H_{S}$ for each cluster in the dendrogram, where results are ordered by colour ranging from blue (lowest value) to red (highest value). The size of each circle represents the flows in each cluster. (b) Plots of the commuting network in the central zone of the country at different thresholds, which correspond to the clusters highlighted in the dendrogram. These plots offer a visual representation of the commuting flows within each cluster and the changes in the diversity patterns and the network structure as the threshold increases.}}
\label{skills}
\end{figure}

In Figure \ref{skills}a, the diversity of skills in each cluster within the dendrogram is depicted, with the cluster containing the capital city exhibiting the highest entropy levels. As the scale increases, smaller city systems with predominantly low entropy and a high dependency on employment opportunities in Santiago join the cluster led by the capital, thus increasing their exposure to diversity. The central-southern area of the country is composed of clusters, that are primarily characterised by medium to high entropy of skills. Figure \ref{skills}b shows how the exposure to skills diversity increases for cities like Chill\'an and Los Angeles as they become part of a larger city system with Concepci\'on at the thresholds of $\tau= 0.16-0.135$. On the other hand, as the urban system that includes Valdivia incorporates smaller and more rural towns, its diversity declines in the following thresholds. Curico maintains a low level of diversity across these thresholds until it joins a larger cluster at the $\tau=0.105$

Examining a city in isolation can limit the overall understanding of its functioning, restrict the identification of diverse skill accessibility, and impede the recognition of opportunities for collaboration to drive economic growth at both local and larger scales.  For instance, planning strategies for a city with a limited range of skills may require different approaches depending on the characteristics of the broader system in which it operates. A city with limited diversity in skills, may have access to a more diverse regional network, offering  a more heterogeneous pool of resources that could potentially supplement the city's own limitations in terms of skills and opportunities. If, on the other hand, this city is part of a region with low diversity or is relatively isolated from other urban centres, it may be necessary to consider other strategies to acquire the skills needed for economic growth. These could include developing strong local businesses and industries that can sustain the local economy, or implementing strategies that foster stronger inter-exchange on a larger scale.

\begin{figure}[h!]
\centering
\makebox[\textwidth][c]{\includegraphics[width=1.1\textwidth]{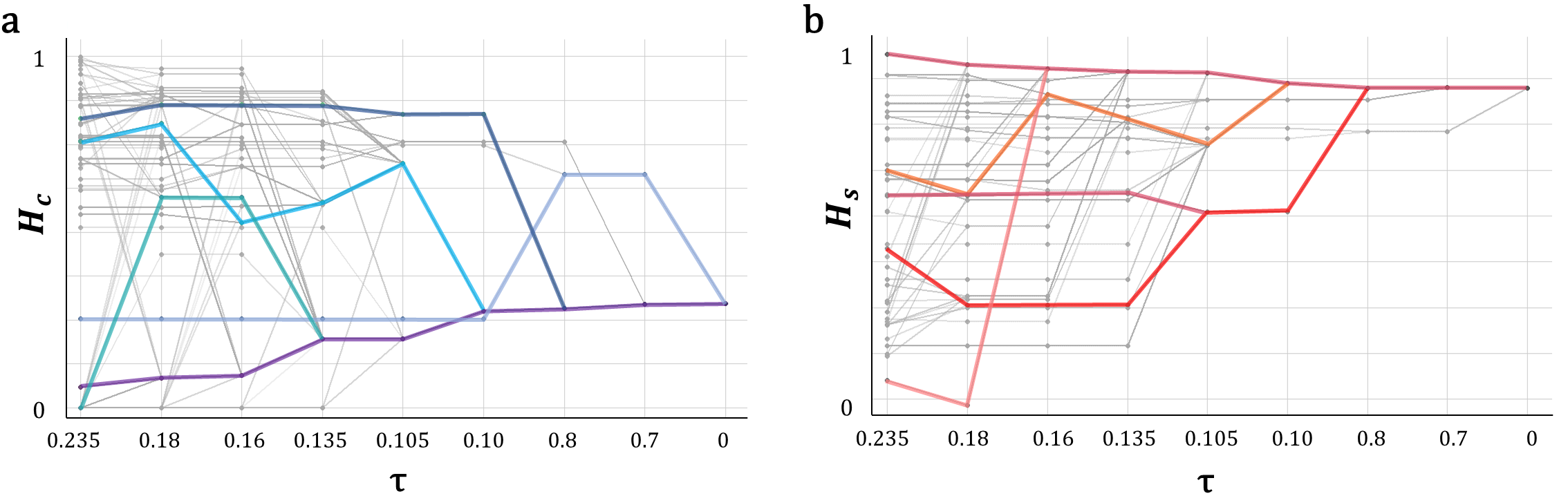}}
\caption{\textit{\textbf{Variation in diversity metrics per city.} The figure displays the diversity values of cities in a given cluster at different percolation thresholds. The x-axis displays the percolation thresholds while the y-axis shows the different diversity values. Each line in the plot represents a city, with highlighted lines added for visual clarity to demonstrate different patterns. a) shows the commuting diversity $H_{C}$, while b) shows the skill diversity $H_{S}$ variation.  }}
\label{lines}
\end{figure}
 
As a summary, Figure \ref{lines}a-b shows the changes in the values of $H_{C}$ and $H_{S}$ for each city as they evolve and merge across thresholds. In both cases, there are cities with very dissimilar patterns, suggesting that the level of exposure to diversity can vary significantly in each case. In some cities, diversity metrics remain stable across thresholds, while in others there are slight or abrupt changes when moving from local to larger scales. Since diversity can vary greatly depending on the scale of observation, planning strategies must be adjusted and refined to address the unique commuting patterns and dynamics of each system of cities across different scales.

\subsection{Systems of cities categories}

\begin{figure}[h!]
\centering
\makebox[\textwidth][c]{\includegraphics[width=1.05\textwidth]{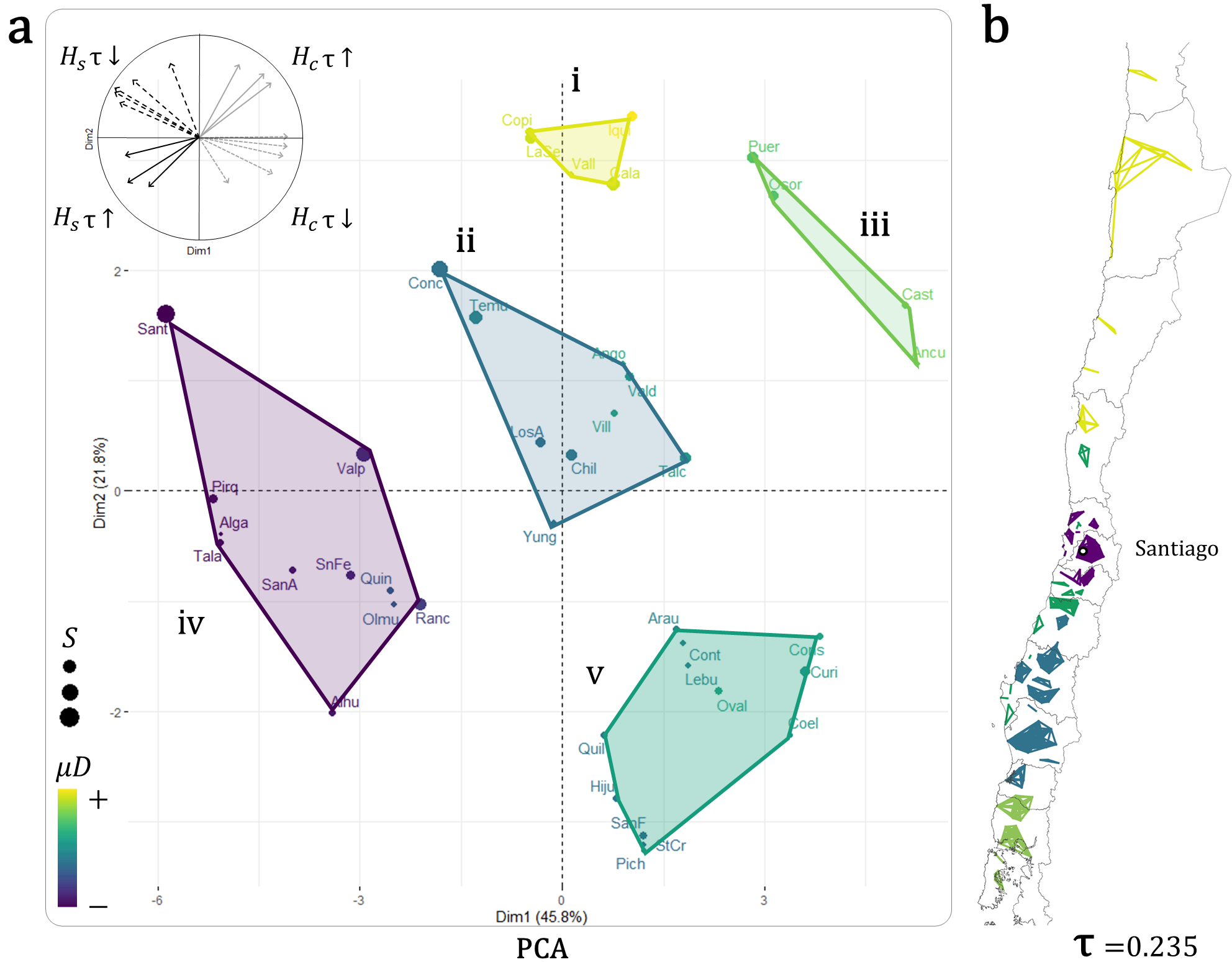}}
\caption{\textit{\textbf{Distinct categories of urban systems based on commuting and skills diversity.} a) Plot of the first two principal components obtained from a PCA analysis using $H_{S}$ and $H_{C}$ values at different scales. The plot includes clusters derived from a k-means analysis of the PCA results. The different categories of cities exhibit distinct behaviours according to the variation of commuting and skills diversity at various scales. The colour represents the mean cophenetic distance of the systems of cities, which shows to be closely related to the resulting categories. In the upper-left corner of the plot, the different variables are displayed, where, for example, $H_{C} \tau\uparrow$ corresponds to a group of variables that belong to the commuting diversity at the upper thresholds. b) Cities coloured based on the different categories obtained from the analysis. }}
\label{methods}
\end{figure}

We conduct a principal component analysis (PCA) using a total of 16 variables, 8 from the commuting diversity $H_{C}$ and 8 from the skills diversity $H_{S}$, each measured at thresholds ranging from $\tau= 0.235$ to $\tau=0.10$, see Figure~\ref{methods}. The two principal components account for 67$\%$ of the total variance in the data. Each point in the plot represents a main city from the systems obtained at threshold $\tau=0.235$. In general, the variables are found to be negatively correlated in each of the quadrants of the plot. Specifically, local and global scales of commuting diversity tend to be negatively correlated with local and global scales of skills diversity. These results suggest that cities with high commuting diversity tend to have lower levels of skills diversity, and vice versa. The PCA outcomes are used for conducting a k-means examination that identified five city clusters displaying unique scalar behaviours and features. Interestingly, cities are arranged in a manner consistent with their average cophenetic distance ($\mu D$), which is further illustrated by the use of ($\mu D$) as a colour code for each city in the plot. This suggests that the degree of dependence between cities at different scales is strongly related to their spatial organisation and, consequently, their skill diversity. The resulting categories can be described as follows:

\textbf{i "Northern urban hubs":} Cities of medium to large size, such as Iquique, Calama, and Copiapo, exhibiting monocentric structures and high skill diversity at a local scale. Despite being geographically isolated, these cities have diversified and potentially harbor self-sustaining economies. At larger scales, they form more polycentric regions with limited intercity exchange. As a result, they have the highest hierarchical distance within the system and a low interdependence at the national level.

\textbf{ii "Regional cities":} Cities with structures that tend to be monocentric at smaller scales but more polycentric at larger scales. They are medium to large-sized cities such us Concepci\'on, Talca and Temuco, and they are well-connected at the national level. Generally, these cities have a diverse economy with small and medium-sized businesses in various sectors, enabling them to act as centres for services and commerce for their regions.

\textbf{iii "Southern specialised centres":} Polycentric city systems with low skill diversity across scales include small to medium-sized cities such as Castro, Osorno, and Puerto Montt, characterised by highly specialised economies based on agriculture, fishing, and livestock. These cities exhibit active local exchange, but limited interdependence with larger systems as shown by high hierarchical distances at the national level.

\textbf{iv "Metropolitan urban systems":} Include two types of systems with the lowest hierarchical distances in the country: those that include major urban centres like Santiago or Valparaiso, and small towns that heavily rely on them, such as Alhue, Quinteros and Talagante. These systems exhibit highly monocentric structures that persist at larger scales due to the dominance of Santiago and Valparaiso throughout the country. Although smaller cities in the system have limited local skill diversity, they gain exposure to greater diversity as they merge with larger cities at the first thresholds.

\textbf{v "In-between cities":} Systems that consist of small cities, like Santa Cruz, San Felipe and Lebu, located in between major urban centres. They exhibit limited skill diversity at the local level and are dominated by a single economic sector, such as forestry, wine production, agriculture, or fishing. Despite being part of a local polycentric system, they are also part of larger monocentric regions, with dominant urban centres, which means they have exposure to a broader range of skills beyond their local area.

\section{Conclusion}

%The study provides new insights on the interrelationships between cities facilitated by commuting, with a particular focus on the inter-dependency that arises from this process. We identified different spatial scales of interaction though percolation analysis, while entropy metrics helped us to better comprehend the interplay between commuting and skill diversity. Our findings are crucial in understanding the patterns of dependence and differentiation that emerge in geographic spaces due to the inter connectivity of cities through daily travel. 

This work establishes a framework to identify inter-dependencies at different geographical scales, that are the result of a heterogeneous availability of jobs and skills provision. The commuting flows resulting from these matching processes, reveal strongly connected cities in certain regions, and weaker relationships in others, forming distinct urban interaction structures with varying skill compositions.

The findings revealed that distinct patterns of spatial interaction can be characterised by diversity metrics based on flow distribution in space and skill diversity levels among workers. We observed that these measures tend to exhibit inverse or negative correlations. This outcome offers alternative perspectives on how spatial distribution and the social composition emerging from urban interaction structures are linked. This may involve adopting a combination of strategies to balance the upsides and drawbacks from different patterns to tackle both economic and social needs. For instance, in a monocentric structure, it may be important to manage the reliance on a single system due to the high concentration of flows, while also benefiting from enhanced social mobility due to diverse human capital. In contrast, a polycentric structure may require capitalising on even regional development while managing limited exchange of human resources due to low skill diversity. Outputs display distinct patterns that vary across scales of observation. Depending on the thresholds, city systems exhibit different urban spatial structures and varying levels of exposure to diversity. Then looking only at a single scale might lead to incomplete or inaccurate conclusions about the dynamics of the system.

These results enable us to identify cities with similar interaction configurations and create categories based on their scalar relationships. Interestingly, we found that these categories are related to the degree of interdependence between cities, as measured by the hierarchical distance between systems. This means that systems with similar overall scalar interactions also share similar patterns of flow spatial distribution and skill composition. This presents a great opportunity for future research to better understand the underlying dynamics of these and other possible relationships, and to further explore the potential of the cophenetic distance as a measure of similarity between urban systems.

While our analysis has focused on daily commuting patterns, including sporadic flows may provide a different perspective on intercity interactions. Additionally, future research could include infrastructure networks, such as transportation and street networks, to examine how spatial accessibility relates to these measures. The significance of this analysis ranges from advancing the quantitative understanding of city systems, to uncovering various types of urban interactions that are relevant for informing strategic political actions. The findings suggest that there is a need and potential for creating new forms of governance and planning that adopt a multi-scalar view and prevent isolating local and regional strategies. Such insights are crucial for unravelling the complex functioning of urban systems and devising effective strategies to foster equitable economic growth and social development.

\bibliographystyle{agsm}

\section*{Acknowledgements}

VM thanks The National Research and Development Agency (ANID) for the financial support provided as a PhD scholarship "Becas Chile".

\section*{Author contributions statement}
VM developed the theoretical framework, designed the experiments and conducted the analysis. CM and EA contributed to the theoretical framework. VM wrote the paper, and all authors contributed to the structure and fine tuning of the manuscript.

\section*{Competing interests}

The authors declare no competing interests.

\end{document}